\documentclass[a4paper,11pt]{article}
\usepackage{pos}
\usepackage{siunitx}
\usepackage{subcaption}

\title{Towards a Study of Low Energy Antiproton Annihilations on Nuclei}
\ShortTitle{Low Energy Antiproton Annihilations on Nuclei}

\author*[a,b]{V.~Kraxberger}
\author[b,c,d]{M.~Bumbar}
\author[e]{A.~Gligorova}

\author[a]{C.~Amsler}
\author[f,g]{M.~Bayo}
\author[h]{H.~Breuker}
\author[a,b]{M.~Cerwenka}
\author[i,j]{G.~Costantini}
\author[f,g]{R.~Ferragut}
\author[g]{M.~Giammarchi}
\author[i,j]{G.~Gosta}
\author[k]{H.~Higaki}
\author[c,d]{E.~D.~Hunter}
\author[a]{C.~Killian}
\author[l]{N.~Kuroda}
\author[i,j]{M.~Leali}
\author[g,m]{G.~Maero}
\author[c,1]{C.~Mal\-bru\-not}
\author[i,j]{V.~Mascagna}
\author[l]{Y.~Matsuda}
\author[i,j]{S.~Migliorati}
\author[a]{D.~J.~Murtagh}
\author[a,b]{A.~Nanda}
\author[a,b]{L.~Nowak}
\author[g,m]{M.~Romé}
\author[a]{M.~C.~Simon}
\author[d,n]{M.~Tajima}
\author[g,m]{V.~Toso}
\author[h]{S.~Ulmer}
\author[i,j]{L.~Venturel\-li}
\author[a,b]{A.~Weiser}
\author[a]{E.~Wid\-mann}
\author[h]{Y.~Yamazaki}

\affiliation[a]{Stefan Meyer Institute, Austrian Academy of Science, Dominikanerbastei 16, Vienna, Austria}
\affiliation[b]{Vienna Doctoral School in Physics, University of Vienna, Boltzmanngasse 5, Vienna, Austria}
\affiliation[c]{Experimental Physics Department, CERN, Esplanade des Particules 1, Geneva, Switzerland}
\affiliation[d]{Faculty of Natural Sciences, Imperial College London, South Kensington Campus, London, UK}
\affiliation[e]{Faculty of Physics, University of Vienna, Boltzmanngasse 5, Vienna, Austria}
\affiliation[f]{Politecnico di Milano, Piazza Leonardo da Vinci 32, Milan, Italy}
\affiliation[g]{INFN Milano, Via Celoria 16, Milan, Italy}
\affiliation[h]{Ulmer Fundamental Symmetries Laboratory, RIKEN, 2-1 Hirosawa, Wako, Saitama, Japan}
\affiliation[i]{Dipart. di Ingegneria dell’Informazione, Università degli Studi di Brescia, via Branze 38, Brescia, Italy}
\affiliation[j]{INFN Pavia, via Agostino Bassi, Pavia, Italy}
\affiliation[k]{Graduate School of Advanced Science and Engineering, Hiroshima University, 1-3-1 Kagamiyama, Higashi-Hiroshima, Hiroshima, Japan}
\affiliation[l]{Institute of Physics, University of Tokyo, 7-3-1 Hongo, Bunkyo-ku, Tokyo, Japan}
\affiliation[m]{Dipart. di Fisica, Università degli Studi di Milano, Via Giovanni Celoria 16, Milan, Italy}
\affiliation[n]{Japan Synchrotron Radiation Research Institute, 1-1-1 Kouto, Sayo-cho, Sayo-gun, Hyogo, Japan}

\note{present address: TRIUMF, Vancouver, Canada}

\emailAdd{viktoria.kraxberger@oeaw.ac.at}

\abstract{
A study of antiproton annihilations at rest on thin solid targets is underway at the ASACUSA facility, which now features a dedicated beam line for slow extraction at \qty{250}{\eV}. The experiment will employ new technologies, such as the Timepix4 ASICs coupled to silicon sensors, to measure the total multiplicity, energy, and angular distribution of various prongs produced in thin solid targets. A detection system consisting of seven Timepix4, covering most of the solid angle, is being constructed. A 3D annihilation vertex reconstruction algorithm from particle tracks in the single-plane detectors has been developed using Monte Carlo simulations. 
The measurements will enable a study of \=p nucleus interactions, 
their dependence on nucleus mass and branching ratios. The results will be used to assess and potentially improve various simulation models.} 

\FullConference{
International Conference on Exotic Atoms and Related Topics and conference on Low Energy Antiprotons\\
EXA/LEAP 2024, 25.-30. August 2024, Vienna, Austria
}

\begin{document}
\maketitle

\section{Introduction}
The antiproton-nucleus (\=pA) annihilation is a key process for studying antimatter and its interaction with matter, as most antimatter experiments use the annihilation products to detect antiprotons or antihydrogen. In annihilations on a single proton or neutron, an average of five pions are produced~\cite{Klempt2005, Amsler1998}, with kaons contributing a few percent. 
The initially produced mesons can trigger a number of final state interactions (FSI) with the residual nucleus, resulting in the emission of heavier particles \cite{Plendl1993}. To date, the characteristics of the FSI have been investigated only for a very small number of nuclei \cite{Bendiscioli1990}. In addition, previous experiments at LEAR (Low Energy Antiproton Ring), along with more recent studies of \=pA annihilation at rest, indicate that current theoretical models struggle to predict all aspects of the annihilation process \cite{Amsler2024, Aghion2017, Egidy1995, Minor1990}. 

The project described here aims to study \=pA annihilations in detail through the detection of both light and heavy charged particles (prongs) in most of the $4\pi$ solid angle. The total prong multiplicity and their angular distribution will be measured for a set of twelve solid targets with thicknesses between 1-\qty{3}{\um}, covering a large range of the periodic table, from carbon to lead.




\section{Experimental Setup}
\begin{figure}[h]
    \centering
    \includegraphics[width=1\linewidth]{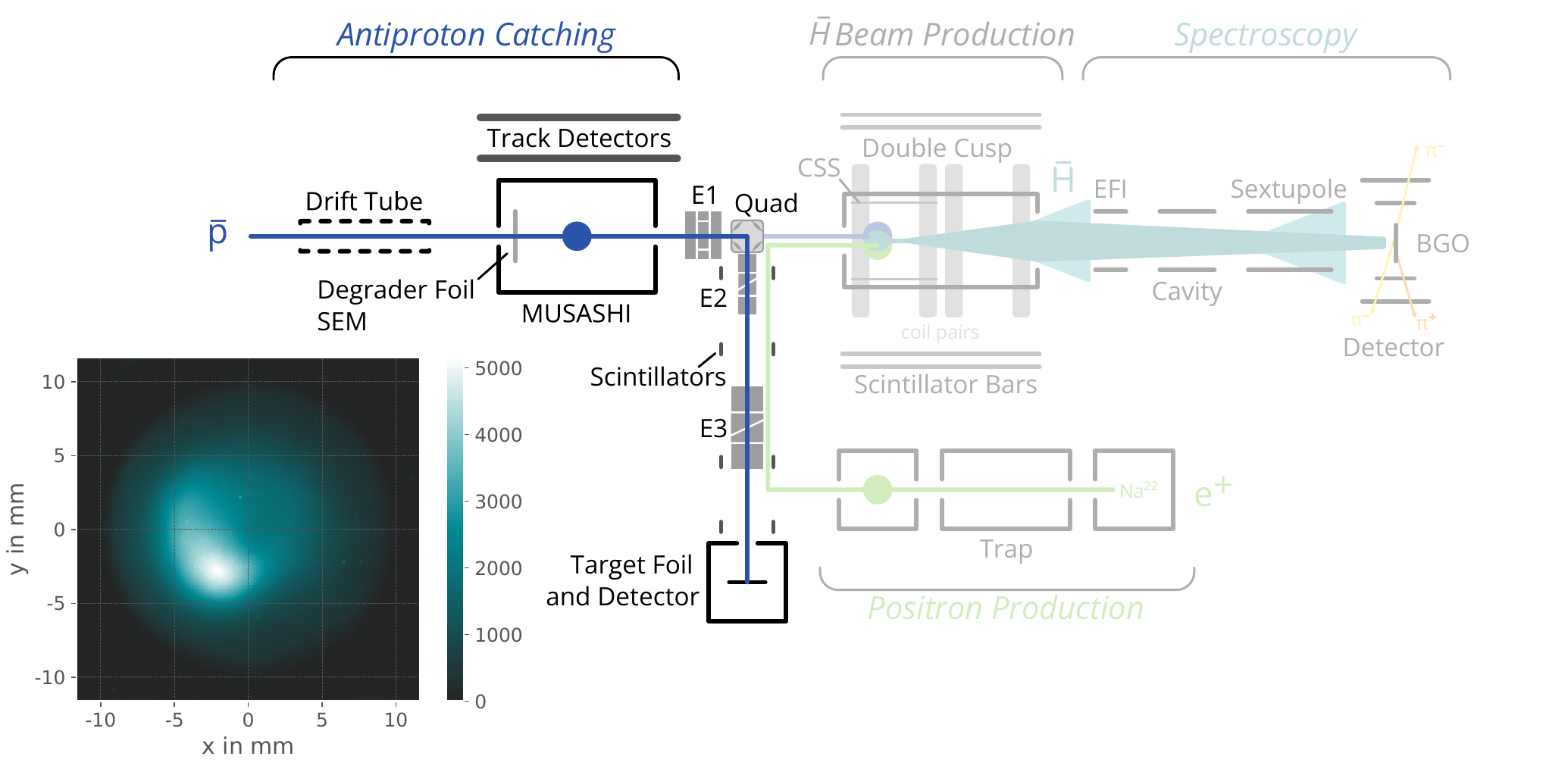}
    \caption{Schematic drawing of the ASACUSA Cusp experiment, with parts not used in this work shown semi-transparent. On the bottom left, an image of \num{\sim 25000} antiprotons annihilating on an MCP (micro-channel plate) is shown at the position of the target foil.} 
    \label{fig:asacusa_setup}
\end{figure}

The primary goal of the ASACUSA Cusp collaboration is to measure the ground-state hyperfine splitting of antihydrogen (\=H)~\cite{Widmann2019}. 
The \=pA annihilation setup was designed as a parallel experiment at the ASACUSA Cusp facility (Fig.~\ref{fig:asacusa_setup}), located at CERN's Antimatter Factory. \\

It consists of two main parts: a beam line for the electrostatic transport of slow extracted antiprotons to the target, and the annihilation chamber which houses the target foil and the detection system. 

\subsection{Antiproton Beam Line}
Antiprotons with a kinetic energy of \qty{5.3}{\MeV} are provided by the Antiproton Decelerator (AD) and are further slowed down to \qty{100}{\keV} in the ELENA (Extra-Low ENergy Antiproton) ring. A pulsed drift tube and degrader foils installed in the ASACUSA apparatus reduce the antiproton energy to a mean value of \qty{\leq 10}{\keV}, enabling their trapping in the MUSASHI (Monoenergetic Ultra-Slow Antiproton Source for High-precision Investigation) trap with an efficiency of ($26 \pm 6$)\unit{\percent} \cite{Amsler2024a}. By lowering the trapping potential, a mono-energetic continuous beam of \qty{250}{\eV} antiprotons can be extracted \cite{Kuroda2012}. The transport of this beam to the annihilation target is made possible by a dedicated electrostatic system~\cite{Bumbar2023}, schematically shown in figure~\ref{fig:asacusa_setup}. After the extraction, an Einzel lens (E1) is used to steer the beam into a quadrupole deflector, in which the beam is bent by \qty{90}{\degree}. The antiprotons are then steered and focused with two Sikler-type Einzel lenses (E2 and E3), minimising the beam size before it reaches the target foil. As shown in Fig.~\ref{fig:asacusa_setup}, the beam line between the deflector and the annihilation chamber is also used for the transport of positrons required for the antihydrogen production. While E1 and E3 were permanently added to the setup and therefore have a diameter large enough not to obstruct the path of the antiprotons and positrons, the deflector and E2 are mounted on a manipulator so that they can be manually inserted when the annihilation experiment is running. 

In 2023, the electrostatic beam line was successfully commissioned using an MCP detector coupled to a phosphor screen and a CCD camera, mounted at the position of the target foil. A maximum of \num{\sim 25000} antiprotons in a beam spot size of~\qty{\sim5}{\mm} FWHM were counted on the MCP during 2~s for every \qty{\sim2}{\min} extraction cycle. The uncorrelated background was estimated to \num{350} counts, based on the counts recorded outside the slow extraction window. The transport efficiency w.r.t. the number of antiprotons that pass through the quadrupole detector, while its electrodes are kept at \qty{0}{\V} potential, is \qty{\sim 8}{\percent}. The majority of beam losses, as indicated by the signals of the scintillators placed along the beam line, occurred at the location of the quadrupole deflector. Since only limited time was spent on the beam optimisation for this result, there is a high probability to achieve a better focus and intensity in the future.



\subsection{Annihilation target and detector}

\begin{figure}[t]
    \centering
    \includegraphics[width=\linewidth]{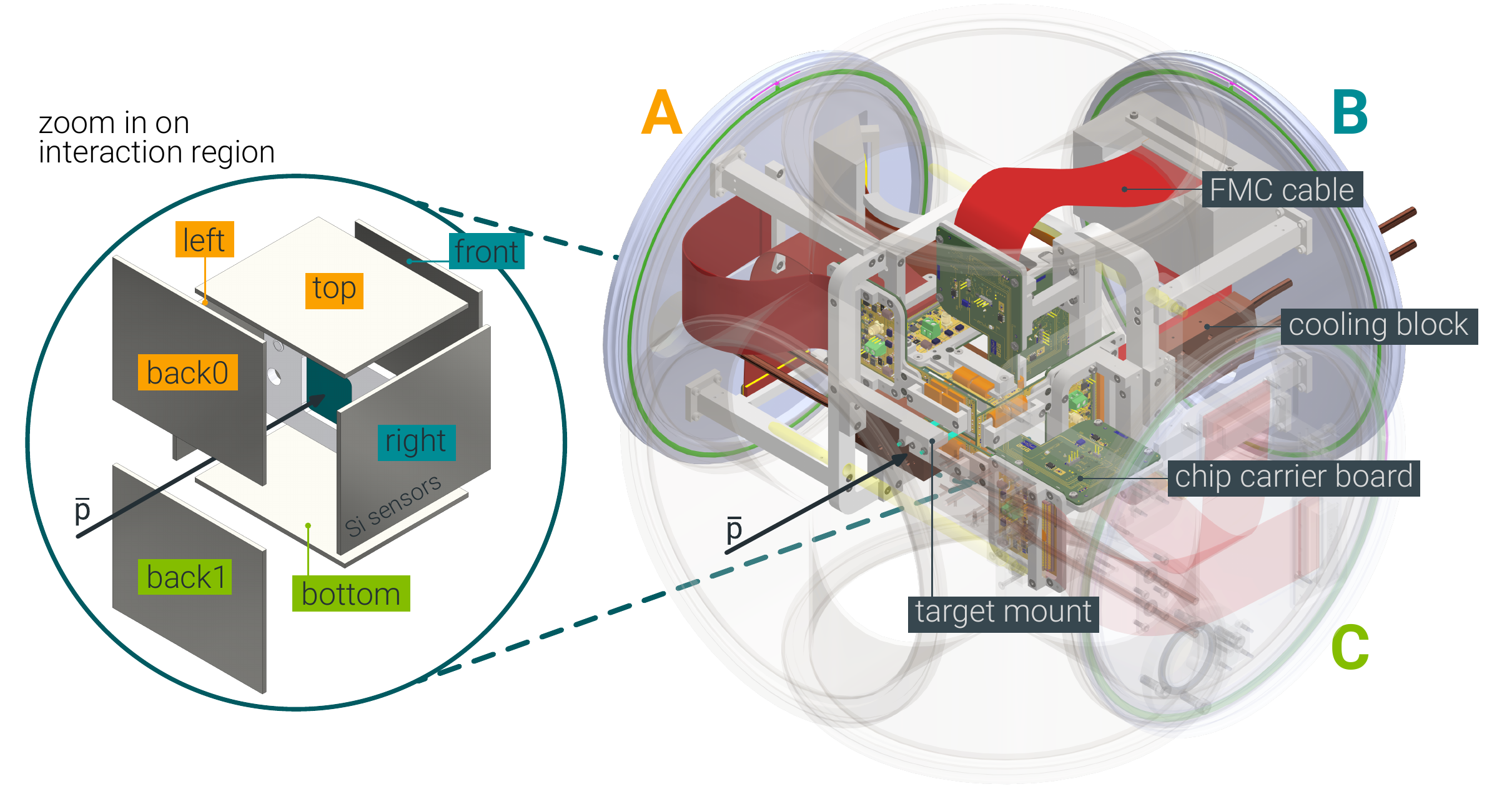}
    \caption{Drawing of the annihilation chamber, where on the right the three flanges A, B and C and labelled in orange, blue and green. On the left a zoom shows only the silicon sensors and the target without the chip carrier boards and mounting structures. The names of the seven sensors are highlighted in the colours of the flange they are mounted on. The direction of the incoming \=p beam is shown with an arrow. The vacuum chamber itself and the flange C are shown with a transparency.}
    \label{fig:chamber_sensors}
\end{figure}

The target foil with area of \qtyproduct[product-units = bracket-power]{1 x 1}{\cm} is surrounded by a detection system consisting of seven Timepix4 ~\cite{Llopart2022} pixel detectors, covering most of the $4\pi$ solid angle. The hybrid pixel readout ASIC (application-specific integrated circuit) consists of $448 \times 512$ pixels of \qty{55}{\um} pitch, resulting in an area of \qtyproduct[product-units = bracket-power]{24.64 x 28.16}{\mm}. Each Timepix4 is coupled to a \qty{500}{\um} thick silicon sensor. The resolution of the time-of-arrival (ToA) is \qty{\sim 200}{\ps}, whereas the time-over-threshold (ToT) energy resolution is \qty{\leq 1}{\keV}. The sensor thickness combined with the good time resolution enable the reconstruction of tracks within the sensor which is described in section \ref{sec:VR}. 
To enable the large solid angle coverage, new chip carrier boards for Timepix4 were specifically designed by NIKHEF, such that the seven sensors can be positioned as tightly as possible in a box-like configuration, shown in Fig. \ref{fig:chamber_sensors}. \\ 
%
%
Three of the six flanges of the vacuum chamber (A, B, C) are used to mount the seven detectors, see Fig. \ref{fig:chamber_sensors}. Flange A holds three detectors as well as the target foil. Flange B, mounted opposite the incoming beam, supports two detectors, while flange C holds another two. To allow for individual installations of A, B and C, each flange is equipped with the necessary feedthroughs for its respective detectors, comprising two or three FMC feedthroughs as well as a Sub-D connector which is used for powering the Timepix4 ASICs and biasing the silicon sensors. 
Since each detector produces up to \qty{8.5}{\W} heat, efficient cooling inside the vacuum is required. To achieve this, each flange has one copper block that is thermally connected to the detectors and cooled using a water-glycol mixture at \qty{10}{\celsius}. With this setup the temperature of the Timepix4 chips remained below \qty{30}{\celsius} while running.
The vacuum poses a major challenge, as the adjacent positron system and MUSASHI require ultra-high vacuum conditions. In the current configuration the pressure inside the annihilation chamber reached the low \qty{e-7}{\milli\bar} level after a few days of pumping.

\section{Monte Carlo Simulations}
The experiment was simulated with Geant4 \cite{Allison2016, Allison2006, Agostinelli2003} and FLUKA \cite{Ahdida2022, Battistoni2015}.
The geometry includes the vacuum chamber, the Timepix4 with the custom boards, the foil holder structure and the foil itself. Due to its complexity, the mounting structure of the detectors was omitted. 
In Geant4 (v11.2.1) the physics list FTFP\_INCLXX\_EMZ was used, which includes the recently added Intranuclear Cascade de Liège (INCL) model \cite{Boudard2013} for antiproton annihilations at rest \cite{Zharenov2023}. The simulated antiproton beam has a kinetic energy of \qty{250}{\eV} in $z$-direction and a Gaussian shape with a standard deviation of \qty{10}{\mm} in $x$ and $y$. In Geant4 the maximum step size of a particle within a medium was set to \qty{5}{\um}. \clearpage

The results of the Monte Carlo simulations 
were digitised using the Allpix$^2$ Semiconductor Detector Monte Carlo Simulation Framework \cite{Spannagel2018}. The energy deposited at different positions was thus converted into induced pixel charge in the biased sensor, matching the signal of the Timepix4 after energy calibration, allowing the same analysis to be applied to both simulation and data without major adjustments.

\section{Vertex Reconstruction}
\label{sec:VR}
\begin{figure}[h]
    \centering
    \raisebox{0.14\height}{\includegraphics[width=0.40\linewidth]{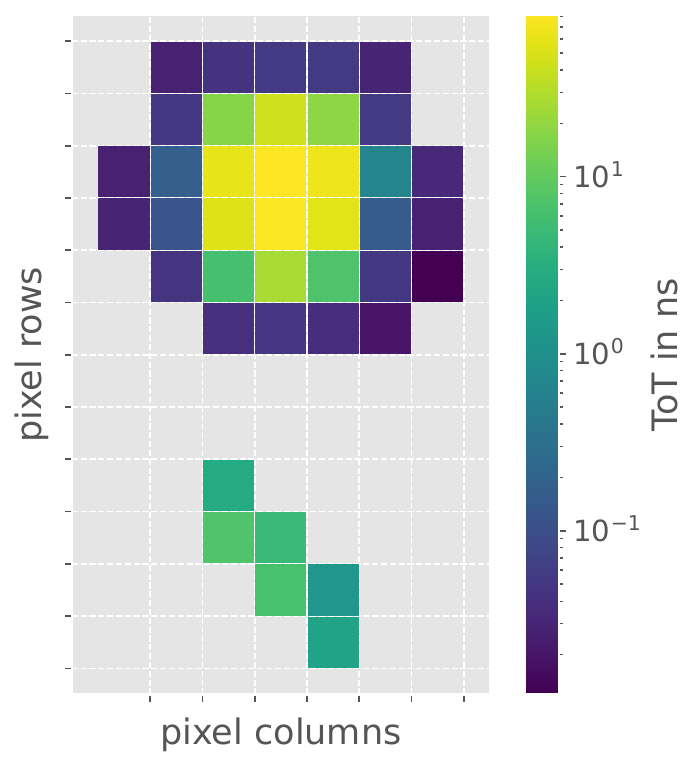}}
    \hfill
    \includegraphics[width=0.5\linewidth]{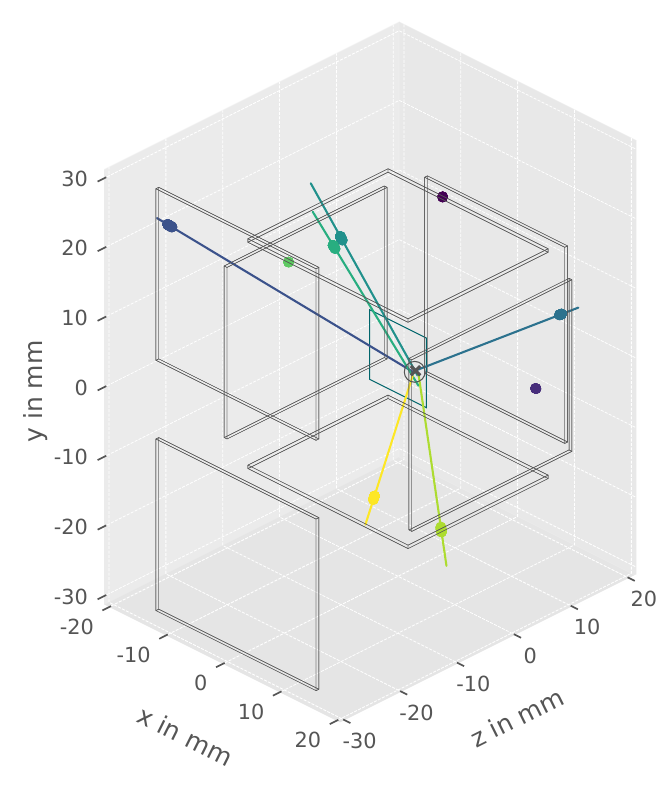}
    \caption{\textbf{Left:} The two typical categories of clusters. On the top a round cluster produced by a HIP, on the bottom a track produced by a MIP. \textbf{Right:} Example of an annihilation event simulated in Fluka. All signals in the Timepix4 detectors are shown as filled dots. If the signal is a track, its linear fit is shown, extracted to the reconstructed vertex (empty circle) on the foil. For comparison the simulated vertex is shown as a cross.}
    \label{fig:rec_event}
\end{figure}

In the upcoming measurements, prongs from one annihilation event will be grouped based on their ToA in the synchronised Timepix4 detectors, and the annihilation vertex will be determined for events with two or more tracks. 
First, the data from each detector is divided into individual clusters using the DBSCAN \cite{Schubert2017} algorithm in python's Scikit-learn \cite{ScikitLearn} package which uses machine learning on the pixel rows, columns and ToA information to combine all signals from a single particle in a cluster. 
While the column and row of a pixel can be used as coordinates, the depth of the particle within the sensor can be extracted from the ToA using several methods \cite{Manek2022}. The minimum and maximum time stamps within a track can be taken as entry and exit points assuming a linear trajectory or the depth can be calculated using the drift time of the charge carriers in the sensor \cite{Bergmann2017}. \\
Minimum ionising particles (MIPs) typically leave a straight track in the detector, while clusters from heavily ionising particles (HIPs) have a more circular shape as they are mostly stopped and deposit all their energy in the sensor. Examples of the two types of clusters are shown in Fig. \ref{fig:rec_event}. Assuming the shape of every cluster to be elliptic, the eccentricity  (focal distance over major axis length) can be used to differentiate tracks from rounder clusters. 
Each track can be fitted and parametrised in 3D with one point on the line and a directional vector. These parameters are then used to find the closest points between all pairs of tracks of one annihilation event. 
The mean of these closest points is taken as the reconstructed annihilation vertex $\vec{V}_\mathrm{rec}$. In Fig. \ref{fig:rec_event} an event simulated in Fluka is shown, where the signals within the sensors are shown as coloured dots, tracks are fitted with lines that are extended to the reconstructed vertex on the foil. $\vec{V}_\mathrm{rec}$ is drawn as an empty circle and for comparison the actual simulated vertex $\vec{V}_\mathrm{sim}$ is shown as a cross. In this event the distance $|\vec{V}_\mathrm{sim} - \vec{V}_\mathrm{rec}|$ is less than \qty{0.8}{\mm}. 

\section{Summary}
A systematic study to explore antiproton annihilations on various nuclei is in preparation at ASACUSA. For this purpose, an electrostatic beam line for slow extracted antiprotons has been constructed, providing approximately \num{25000} antiprotons at the target position during 2~s every \num{\sim2}~min. A detection system with seven Timepix4 ASICs coupled to \qty{500}{\um} silicon sensors will be used to measure the multiplicities of MIPs and HIPs and their angular distributions for a broad range of thin target foils. A vertex reconstruction algorithm has been developed using Monte Carlo simulations which enables the tagging of individual annihilation events in future measurements. 

\newpage
\section{Acknowledgments}
This work is supported by JSPS KAKENHI Fostering Joint International Research No. B 19KK0075, No. A 20KK0305, Grant-in-Aid for Scientific Research No. B 20H01930, No. A 20H00150; the Austrian Science Fund (FWF) Grant Nos. P 32468, W1252-N27; Special Research Projects for Basic Science of RIKEN; Università di Brescia and Istituto Nazionale di Fisica Nucleare; the European Union’s Horizon 2020 research and innovation program under the Marie Skłodowska-Curie Grant Agreement No. 721559; the Research Grants Program of the Royal Society and the Foundation for Dutch Scientific Research Institutes. The authors express their gratitude to the Medipix4 Collaboration at CERN for providing the Timepix4 ASICs for the measurements, NIKHEF for providing the custom designed chip carrier boards, in particular to Martin van Beuzekom, and to the late Hannes Zmeskal for his substantial contribution in the design of the experiment.
This research was funded in whole or in part by the Austrian Science Fund (FWF) P 34438. For open access purposes, the author has applied a CC BY public copyright license to any author accepted manuscript version arising from this submission.

\newpage
\bibliographystyle{JHEP}
\bibliography{references}


\end{document}